# Influence of Collimation and Detector Length on CT Exposures Measured in a 60 cm Long Body Phantom

Victor J Weir

## ABSTRACT

**Background:** Approach to equilibrium functions are typically presented as growth functions where the exponential term contains scan length as a variable. Data fitting approaches to modeling allow for parametric formulae that my not reveal the physical quantities involved and their influence on the function being studied. In this paper the author models approach to equilibrium function by a method that allows the physical quantities to be defined beforehand, allows their influence on the model to be studied, and can be used to predict how each physical quantity affects approach to equilibrium. **Methods:** An ordinary differential equation (ODE) is used to model the approach to equilibrium function. Two equivalent ODEs are derived and one is used for modeling the approach to equilibrium function for the case where collimation is changed at fixed detector size. A parallel model is used to study the approach to equilibrium as a function of detector size where collimation is fixed. Both models are validated by experimental measurements in a 60 cm body phantom. Influence of detector size is simulated by using leaded sleeves of varying sizes wrapped around a 100 mm pencil chamber. This creates sleeve gaps of 10 – 60 mm around the chamber. A 100 mm detector size is represented by the pencil chamber with no leaded sleeves. **Results:** The model accurately describes experimental data for the cases studied. Except for the smallest collimation of 5 mm, linear regression analysis shows good correlation between data and model according to $R^2$ values at all collimations. Further analysis of the model reveals the minimum detector size that will integrate 98% of the signal for a given collimation is $d = 4.94L$. Additional analysis shows the minimum collimation at which a given detector will collect 98% of the signal is $L = 5.49d$. **Conclusion:** Modeling by differential equations is an accurate method for the approach to equilibrium function and has advantages over typical curve fitting methods.

# INTRODUCTION

A model is a way to explain or describe physical data. The usefulness of the model is in its ability to predict future physical results with some degree of accuracy. When exposures are measured in a CT phantom of "infinite" length and the exposures approach equilibrium at longer scan lengths it suggests that the detector is not collecting all the available exposure. The detector reaches an asymptote measurement and will return the same measurement value at ever widening beam widths. One way to solve this problem is to use a longer detector in an "infinite" phantom or to use a "point" detector in an "infinite" phantom and to assume that the peak dose measured by the "point" detector is all that is needed to describe dose. The approach to equilibrium can be described by a mathematical growth model. The majority of models utilize the exponential growth function. Growth functions may be of the logistic (Verhulst) type described in [1] or of the exponential type such as Gompertz [2], or Weibull type described in [3]. A generalized growth function was derived by Koya-Goshu [4]. Our model leads to a growth function of the Gompertz [2] type.

A previous paper from Weir [5] defined and stated a necessary and sufficient condition for the convergence of the dose profile at infinite distances from the location of the detector. If the dose profile is shown to converge in a long enough phantom, we can say that signals from the portions of the dose profile far from the detector will contribute no exposure to the detector at its location. This will imply that if a detector is placed at a location and several scans are taken at different increasing collimations the signal measured by the detector will approach an equilibrium value assuming the phantom is infinitely long. The analogous idea is to use the scan length as the variable in the approach to equilibrium [6] or to scan an infinite phantom (900 mm length) at different beam widths [7]. Others have used Monte Carlo simulations to demonstrate the approach to equilibrium based on the scan length as a variable [8-10].

In the papers using Monte Carlo simulations [10], no mention was made of the influence of detector size or of the relationship between collimation length and detector size on the approach to equilibrium. It is logical to expect that when a longer detector is used for measuring signals, the detector will measure signals that are at farther distances from it compared to a smaller detector, provided we fix efficiencies, measurement volumes, sensitivities, etc. The smaller detector should approach equilibrium faster than the larger detector under the same conditions. Furthermore, we can expect that the collimation of the beam will play a role in how soon a detector reaches equilibrium when detectors of different dimensions are compared. Nevertheless, the explicit influence of collimation and detector dimension has not been explored in previous works until now. In this paper we develop and present a mathematical model that shows how the approach to equilibrium of peak exposures (approach to equilibrium) measured by a detector depends on the dimensions of the detector as well as collimation set. This is done under the condition of convergence of the dose profile, achieved by using a body phantom of 60 cm length. We present several experimental measurements to validate the model and analyze various scenarios that determine the minimum collimation needed for a detector of a given size to reach a steady state (i.e., approach equilibrium). In other words, we can determine the collimation at which a detector begins to act as a point detector. A secondary result is also the length of detector needed to integrate 98% of the dose profile at a given collimation. We present an innovative approach to simulate integrating detectors of varying size by utilizing a partial volume effect on the long pencil chamber. We accomplish this by the use of a leaded sleeves around the pencil chamber. With this design, we can simulate the approach to equilibrium when the limits of integration are changed (changing the sleeve gap). We also simulate the approach to equilibrium when the limits of integration are fixed (fixed sleeve gap) and the beam collimation is increased. Furthermore, data collected from a Farmer chamber can be shown to be identical to the data measured by a sleeve gap of length equal to the length of the Farmer chamber.

This work is different from the use of continuous lead sleeves as well as the use of sleeves with apertures. The use of continuous lead sleeves and lead sleeves with apertures have been used previously to measure exposure sensitivity profile [11]. In that work it was concluded that leaded sleeves with apertures could be used to measure peak exposures from single scans.

# MATERIALS AND METHODS

## Description of models

We develop a mathematical model leading to the exponential growth function shown below.

$$D(z) = D_0 \left(1 - e^{-\frac{\alpha z}{d}}\right)$$

Others have used versions of the exponential growth function to model approach to equilibrium. In these models, the term that multiplies the parenthesis is typically called a scatter fraction and d was defined as the width of the line spread function (LSF). In both [6] and [12], d = 117 mm. In another paper, d was termed a fitting parameter and ranged from ~65 - ~120 mm [8] and ~84.5 – 129 mm [13]. Those papers arrive at their model by data fitting approaches or by Monte Carlo analysis [10]. Data fitting unfortunately does not allow one to identify the actual parameters in the model and may lead to models that do not fully explain the problem at hand. In this paper, a model is developed based on the use of differential equations.

## The model

Consider that peak weighted exposure $y$ depends on the collimation L and the length d of the integrating detector used for collecting the exposure. Assuming that the detector is placed in a phantom of infinite length, we can say the dose profile converges to zero at infinite distances from the detector. We define peak exposure as the exposure measured when a detector is placed at z=0 in the central and peripheral axis locations, and when the dose profile converges to zero at the tail ends of the phantom. We consider "converges to zero" to mean that the signal at the tail ends is less than 2% of the dose profile.

In this paper, a model is developed that leads to an ordinary differential equation (ODE). The model is solved with initial conditions set and integrating constants determined.

We start by modeling the approach to equilibrium of peak exposures. We can identify collimation and detector dimension as important independent variables. Weighted exposure (exposure) is an important dependent variable described by a real valued function of a single real variable $y = y(z)$. Let $y(z): R \to R$ be a continuously differentiable function and $z \in R$. Then $y(z)$ is $C^\infty(R)$ in the Euclidean space $R$. It is also analytic about some point $z_0$ in the domain, where $z_0 \in R$. Although scan length $(S_L)$ is not explicitly modeled in this paper, it is worth noting that scan length is related to collimation as a real number multiple of the collimation (i.e., $S_L \propto L$). Table movement per rotation is equal to pitch multiplied by collimation. Scan time (s) multiplied by table movement in mm/s determines the scan length (mm). We assume the phantom is infinitely long and thus all exposure profiles converge at the tail ends of the phantom. We are led to consider two questions - (1) how does exposure change with collimation for different sized integrating detectors? and (2) how is the result related to the current 100 mm pencil chamber and the 23.5 mm Farmer chamber? To answer these questions we can model changes in weighted exposure $y(z)$ with collimation as the solution of the following ODE for the case where we keep the detector d fixed (sleeve gap fixed) and collimation is changed. Collimation is initially represented by the z variable. The following differential equation is used as a model:

$$\frac{d}{\alpha} y'' + y' = 0, z \in [0, L]$$

where L represents the available collimations on the CT scanner and goes from 5 mm – 160 mm. $y(z)$ is the weighted exposure at each collimation z. $y(z)$ is $C^\infty(R)$ in the Euclidean space. d is the set sleeve gap and is a surrogate for detector dimension for an integrating detector, and α is an undefined shaping parameter.

A general solution of this equation is

$$y(z) = \frac{d}{\alpha} k_1 e^{-\frac{\alpha z}{d}} + k_2$$

where $k_1$ and $k_2$ are integration constants. Note that if we allow $k_1$ to be zero then $y(z) = k_2$ is a trivial solution of the above ODE. The initial conditions are
$$y(0) = H_{min} \geq 0, \ y'(0) = B$$

The initial measurement $H_{min}$ is either zero or close to zero. B is an undefined scaling factor.

Under these initial conditions, $k_1 = -B$ and $k_2 = H_{min} + B\frac{d}{\alpha}$.

After a little substitution and algebra, we arrive at the equation for our model:

$$y(z) = B\frac{d}{\alpha}\left(1 - e^{-\frac{\alpha z}{d}}\right) + H_{min}$$

If we instead take the initial conditions $y(0) = H_{min} = 0$, then

$$y(z) = B\frac{d}{\alpha}(1 - e^{-\frac{\alpha z}{d}}) \tag{1}$$

where the term $B\frac{d}{\alpha}$ represents an asymptote and is a theoretical maximum value for y that depends on the sleeve gap used for measurements in this case. The model above is a derivative of the first order ODE shown below.

$$\frac{d}{\alpha}y' + y - H = 0 \ z \in [0,L]$$

$y(z)$ and d have been previously defined above. H is a critical value, an asymptote, in theory a maximum value for $y(z)$ at a given collimation. Again α is an undefined shaping parameter. A general solution of this equation is

$$y(z) = k_1 e^{-\frac{\alpha z}{d}} + H$$

where $k_1$ is an integration constant. If we allow $k_1 = 0$, then
$y(z) = H$ is a trivial solution of the ODE. The initial condition is $y(0) = H_{min}$. Under these initial conditions

$$k_1 = H_{min} - H$$

After a substitution and algebra we arrive at the equation for our model:

$$y(z) = H\left(1 - e^{-\frac{\alpha z}{d}}\right) + H_{min} e^{-\frac{\alpha z}{d}}$$

Both of these models are validated by experimental measurements in a 60 cm long body phantom. The $H_{min}$ term is the initial reading at the smallest collimation and is therefore expressed as the minimum exposure. H is an asymptotic term and is a theoretical maximum for $y(z)$. If $H_{min}$ is taken to be zero in the initial condition,

$$y(z) = H\left(1 - e^{-\frac{\alpha z}{d}}\right) \quad (2)$$

Equations 1 and 2 are identical except for constant factors outside the parenthesis. Suppose we reverse the influence of collimation and detector dimensions, how does the exposure measured by an integrating detector change? The model above can be recast to study how the exposure changes at different detector sleeve gaps d for a given collimations L. The sleeve gap is used as a surrogate for detector dimension so as to simulate detectors of different integrating lengths. In this case we represent the sleeve gap by the variable x and collimation by L. We can rewrite the ODE as

$$\frac{L}{\alpha} y'' + y' = 0 \quad z \in [0, d]$$

where d represents the sleeve gaps used, and goes from 10 mm – 100 mm.

The above is the derivative of the differential equation

$$\frac{L}{\alpha} y' + y - H = 0$$

The results of these ODEs follow those above and will not be repeated here. We will just state briefly that the solution for the equation

$$\frac{L}{\alpha} y'' + y' = 0 \quad z \in [0, d]$$

is like equation 1 with the sleeve gap d replaced by the collimation L. Likewise, the solution for the equation

$$\frac{L}{\alpha} y' + y - H = 0$$

Is like equation 2 with sleeve gap d replaced by collimation L.

Based on the preceding analysis we can come up with the following definition

**Definition:** Peak exposure in CT is the exposure measured at any collimation by a detector whose dimension d approaches a point size i.e. $d \to 0$.

**Theorem:** If the detector dimension $d \to 0$, then the measured exposure $y(z)$ approaches the peak exposure $H$, i.e. $y(z) \to H$.

**Proof:** Using equation 2,

$$y(z) = H\left(1 - e^{-\frac{\alpha z}{d}}\right)$$

$$\lim_{d \to 0} y(z) = \lim_{d \to 0} H\left(1 - e^{-\frac{\alpha z}{d}}\right)$$

Since $e^{-\frac{\alpha z}{d}} \to 0$ when $d \to 0$ at any $z > 0$, we have $y(z) \to H$. ∎

## Experimental validation

A validation of this model was done as follows. Four body phantoms were assembled as shown in Figure 0(A) and used for measurements in a 256 slice GE Revolution CT scanner at our institution. The scanner had collimations of 5 mm, 40 mm, 80 mm, 120 mm, 140 mm, and 160 mm. All scans were done at 120 kVp, 120 mAs (1 second rotation), and in axial mode. This technique was chosen since previous work has shown that the dose profile converges to zero at the tail ends of the 60cm body phantom for the largest collimation available on this CT scanner [5].

Exposure was collected by a 100 mm active length model 10x5-3CT pencil chamber connected to a recently calibrated Radcal Electrometer (Radal Corporation, Monrovia CA). The pencil chamber is known to exhibit excellent positional uniformity and partial volume response. The plastic sleeve of the pencil chamber was removed and the lead sleeves placed over the chamber (Figure 0(B)). The chamber was then inserted into the holes in both the central and peripheral locations, z = 0, during measurements. To check that the lead sleeves blocked all radiation, an initial measurement of the pencil chamber with the entire active length covered by a lead sleeve was done at 40 mm collimation. Exposure, if any was measured, was called $H_{min}$. Lead sleeves were measured with a pair of Vernier calipers and found to be on average 1.5 mm thick. Eight leaded sleeves constructed of lengths 5 mm, 10 mm, 15 mm, 20 mm (2), 40 mm (1), and 60 mm (2). These were placed in different arrangements to create sleeve gaps of 10, 20, 30, 40, 50, and 60 mm. When there was no sleeve on the pencil chamber, that was considered to be a 100 mm gap. Exposures were weighted as 0.333*central + 0.667*periphery.

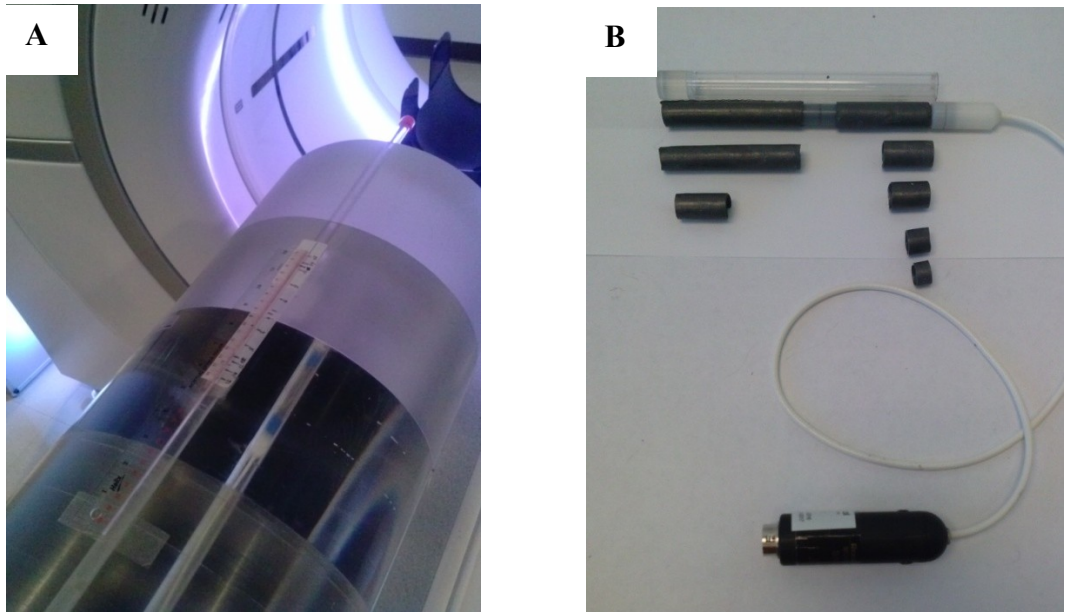

Figure 0. (A) Picture of the 60 cm long body phantom and (B) picture of the pencil chamber with various lead sleeves used for creating the sleeve gaps.

## Measurement of exposure at different collimations for a fixed sleeve gap

The first experiment was to vary collimation for a fixed sleeve gap. All measurements were done with the sleeve gaps centered on the active volume of the chamber. A technique of 120 kVp and 120 mA (1 sec rotation) was used for all measurements. At each fixed sleeve gap, exposures were taken at different collimations. Measurements were repeated at other sleeve gaps for all available collimations. Plots of results are in Figure 1.

## Measurement of exposure at different sleeve gaps for a fixed collimation

In this experiment the collimation was fixed in a 60 cm long body phantom, and the sleeve gap was varied by using different sleeve combinations. All measurements were done with the sleeve gaps centered on the active volume of the chamber. For each new collimation set, measurements of exposure were done at different sleeve gaps. A graph of the results is presented in Figure 2.

## Prediction of minimum collimations at different detector dimensions

Weighted exposure data was normalized to the smallest collimation and plotted against the normalized collimation and sleeve gaps to produce Figure 3. This was analyzed to predict the minimum collimation at which a detector behaves as a point detector. We did this by assuming that in an infinite phantom, the dose profile is converged when the tail ends represents 2% of the signal. Thus 2% was used as a cutoff value for predicting the relationship between collimation and detector dimension in an infinite phantom. Results are summarized in Table 1.

## Prediction of minimum detector dimensions at different collimations

Another interesting feature of our results is the ability to determine the detector length that will integrate 98% of the exposure signal. Weighted exposure data was normalized to the smallest sleeve gap and plotted against the normalized dimension represented by the ratio of sleeve gap to collimation. Results are plotted in Figure 4. A summary of this is also in Table 2.

## Model fitting procedures

Data in Figures 1 and 2 are fitted with our model and analyzed to determine goodness of fit. Each data point in Figure 1 was fit by varying the parameter α and the H term in front of the parenthesis. The model was improved by minimizing the sum of squared residuals (RES) using solver in Excel (Microsoft, Bellevue, WA). This was accomplished by minimizing the squared difference between the actual experimental values $y_i$ vs. the modeled values $\hat{y}_i$.

$$RES = y_i - \hat{y}_i$$

This was used as a starting point to adjust the parameters for α and H for the best fit and most consistent result. These parameters are reported in Tables 1 and 2 for the various scenarios studied. For goodness of fit, we did a linear regression analysis of the actual and experimental data points. Additional model evaluation was done by calculating the Root Mean Square Error (RMSE) by

$$RMSE = SD_{y_i} * \sqrt{1 - R^2}$$

The results are shown in Table 3.

# RESULTS

## Exposure at different collimations for a fixed sleeve gap

Below is the experimental measurement of the change in exposure with the collimation for lead sleeve gaps around the pencil chamber. Measurements are taken with a 100 mm pencil chamber. The graphs are modeled by the equation

$$y(L) = H\left(1 - e^{-\frac{\alpha L}{d}}\right) + H_{min} e^{-\frac{\alpha L}{d}} \qquad (3)$$

where α is a shape parameter used for improving the fits and H is related to detector dimension and collimation. $H_{min}$ is the minimum exposure at zero collimation. An $H_{min}$ of between 12 mR and 15 mR was used to fit graphs in Figure 1. We can also define a length constant $\beta = \frac{d}{\alpha}$, where $\beta$ has the unit of a dimension (mm).

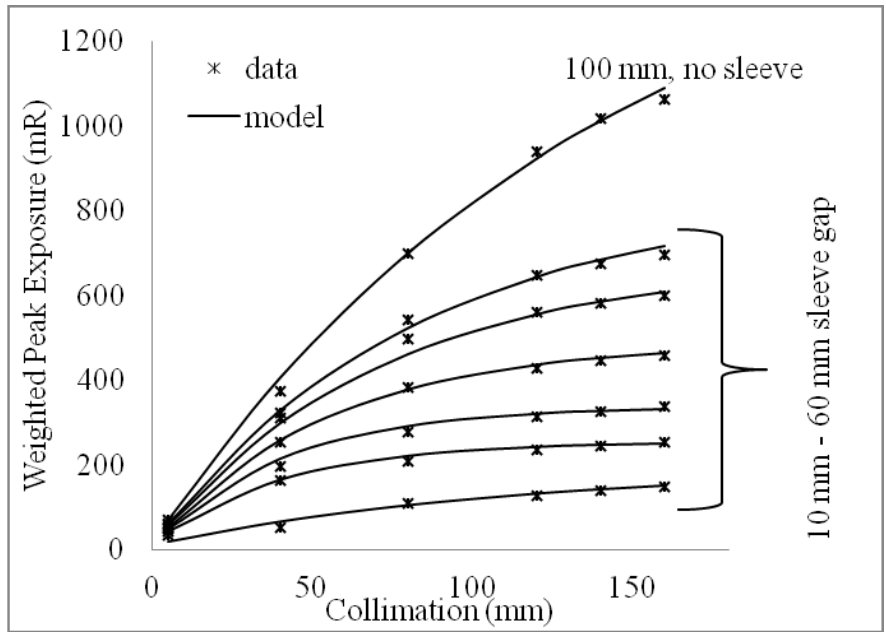

Figure 1. Weighted exposure plotted against the collimation for a 256-slice GE Revolution CT scanner. The exposure at each collimation is collected at different sleeve gaps from 10 mm – 60 mm and then at 100 mm (pencil chamber with no plastic cover).

# Exposure at different sleeve gaps for a fixed collimation

Below is the experimental measurement of the change in exposure with the sleeve gaps for each collimation. Measurements are taken with lead sleeves around a 100 mm pencil chamber. The data is modeled by the equation

$$y(d) = H\left(1 - e^{-\frac{\alpha d}{L}}\right) + H_{min} e^{-\frac{\alpha d}{L}} \quad (4)$$

where α is a shape parameter used for improving the fits and H is related to collimation and detector dimension. An $H_{min}$ of about 15 mR was used to fit graphs in Figure 2.

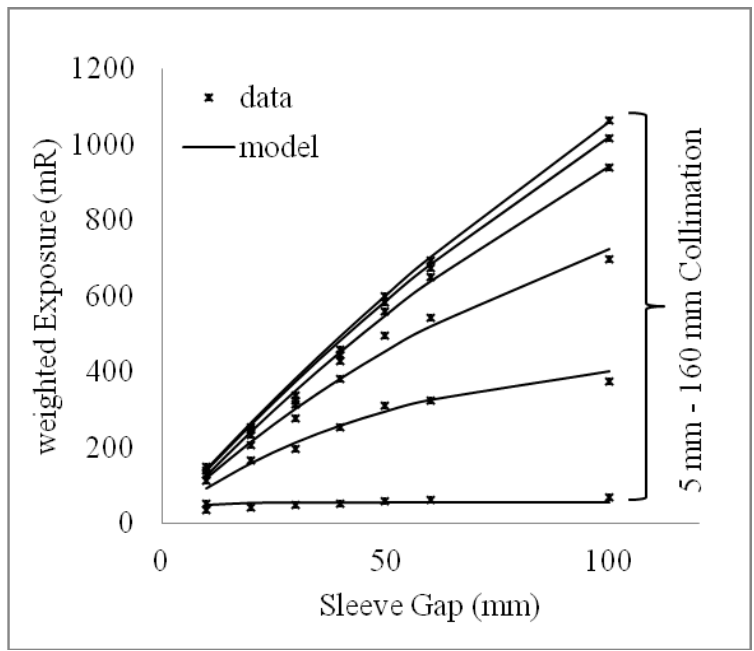

Figure 2. Weighted exposure plotted against sleeve gap (mm) for a 256 GE Revolution scanner. The exposure at each sleeve gap is collected at different collimations of 5mm, 40mm, 80mm, 120mm, 140mm, and 160mm. Highest plot is at 160mm collimation, lowest plot is at 5mm collimation.

## Prediction of minimum collimations at different detector dimensions

To predict minimum collimations at different detector dimensions, we used the data for plotting Figure 1. The data was re-plotted with respect to the ratio of collimation and detector dimension. This graph is shown below in Figure 3.

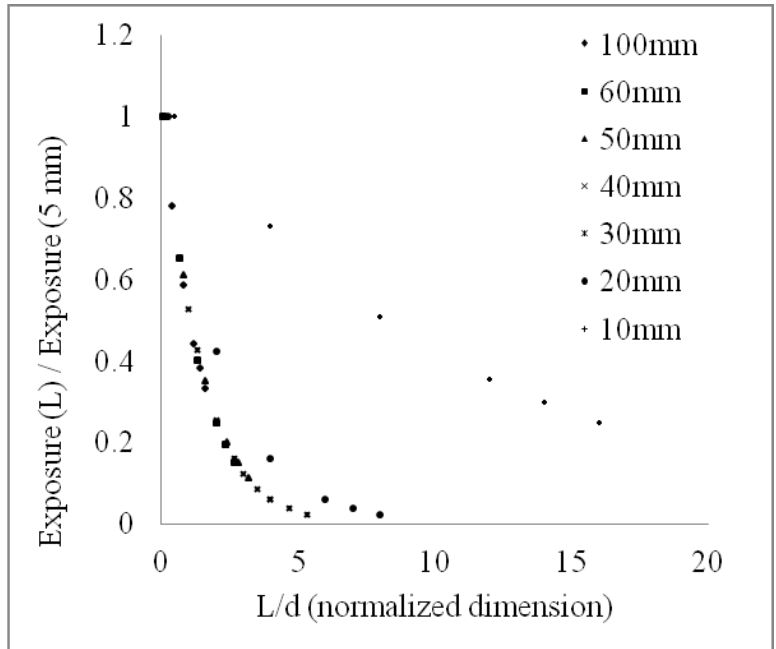

Figure 3. Ratio of weighted peak exposures at different collimations relative to the smallest collimation setting (5 mm) is plotted against a normalized dimension involving the ratio of collimation L to detector dimension d. The curve for all sleeve gaps from 10 mm to 100 mm is shown for a 256-slice GE Revolution CT scanner.

If we consider the variation of exposure with collimation at fixed sleeve gap, we have for the 100 mm sleeve gap

$$\frac{Exp(L)}{Exp(0)} = e^{-\frac{0.71L}{d}}$$

$$L = 5.49d \qquad (5)$$

Table 1 below is a complete table of possible collimations L for the different sleeve gaps d used.

Table 1. Fitting parameters are shown at different sleeve gaps (surrogate detector dimension). The collimation (L) at which an integrating detector of dimension equal to the sleeve gap will behave as a point detector is also shown.

| Sleeve gap d (mm) | α | H(mR) | Collimation L (mm) |
|---|---|---|---|
| 10 | 0.09 | 195 | L=43.44d |
| 20 | 0.49 | 256 | L=7.98d |
| 30 | 0.73 | 340 | L=5.36d |
| 40 | 0.73 | 490 | L=5.36d |
| 50 | 0.70 | 680 | L=5.59d |
| 60 | 0.73 | 835 | L=5.36d |
| 100 | 0.71 | 1600 | L=5.49d |

## Prediction of minimum detector dimensions at different collimations

To predict minimum detector lengths at different collimations we used the data for plotting Figure 2. The data was re-plotted with respect to the ratio of detector dimension and collimation. The results are shown below in Figure 4.

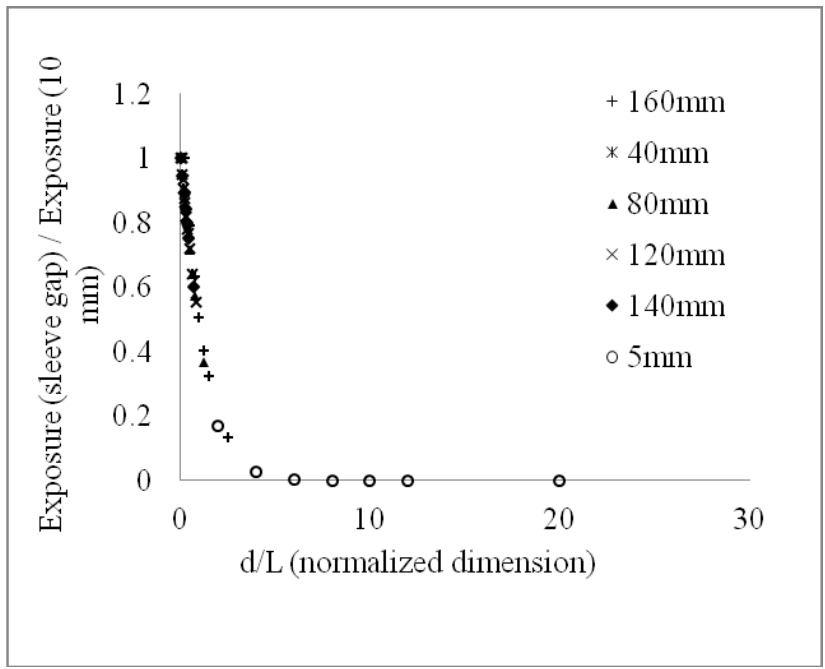

Figure 4. Ratio of weighted peak exposures at different sleeve gaps to the smallest sleeve gap setting (10 mm) is plotted against a normalized length factor involving the ratio of detector dimension (d) and collimation (L). The curve for all collimations from 5 mm to 160 mm is shown for a 256-slice GE Revolution CT scanner.

If we take 2% to be the cutoff point for stable equilibrium to be achieved, then using

$$\frac{Exp(d)}{Exp(0)} = e^{-\frac{0.79d}{L}}$$

we see that for variation of exposure with sleeve gap (d) at fixed collimation (L),

$$d = 4.94L \qquad (6)$$

A complete table of the possible detector lengths is shown below in Table 2.

Table 2. For each collimation set in the first column, the fitting parameter α is shown in the second column. The fourth column shows the length of sleeved chamber needed to collect 98% of the signal. This is for data collected in a 60 cm long body phantom at both central and peripheral positions using the technique factors stated earlier.

| Collimation L(mm) | α | H(mR) | Minimum detector length d (mm)* |
|---|---|---|---|
| 5 | 0.89 | 55 | 21.96 |
| 40 | 0.79 | 465 | 197.97 |
| 80 | 0.79 | 1145 | 395.95 |
| 120 | 0.79 | 1950 | 593.92 |
| 140 | 0.79 | 2350 | 692.91 |
| 160 | 0.79 | 2700 | 791.89 |

*Note that the detector length also sets the minimum phantom length needed for the dose profile to converge at the tail ends of the phantom. In this case we considered 2% to be the cutoff for convergence.

## Model fitting results

Table 3. Evaluation of the model by linear regression. Experimental $(y)$ vs. model predicted $(x)$ exposures and their $R^2$ value. Additional model evaluation is by the root mean square error RMSE.

| Exposure at different sleeve gaps for a fixed collimation | | | |
|---|---|---|---|
| Collimations | Linear regression | $R^2$ | RMSE |
| 5 | y=0.1764x+44.448 | 0.5625 | 7.55 |
| 40 | y=1.0249x | 0.9606 | 21.89 |
| 80 | y=0.993x | 0.9848 | 25.36 |
| 120 | y=1.0069x | 0.9949 | 19.70 |
| 140 | y=1.0217x | 0.9935 | 24.02 |
| 160 | y=1.0148x | 0.9947 | 22.55 |

# DISCUSSION

Figure 1 shows experimental data for the exposure at different collimations for a fixed sleeve gap. After data collection at each collimation, the sleeve gap was changed and data collected at that sleeve gap for different collimations. The smallest sleeve gap used was 10 mm. Due to the magnitude of the data collected at the 10 mm sleeve gap, there is likely more error at such a small sleeve gap compared to the larger sleeve gaps. Any penetration at such small sleeve gaps will have a larger impact. At 20 and 30 mm sleeve gaps, the line reaches an asymptote and levels off as expected. This can be seen for the 40 mm sleeve gap to some extent. At 60 and 100 mm sleeve gaps, the asymptote is not reached and the lines continue to grow indicating that the asymptote is at a larger collimation than is available on the scanner. The model, Equation 3, appears to slightly overestimate at larger sleeve gaps as can be seen at the peaks of the plots in Figure 1 for the 50 mm, 60 mm, and 100 mm sleeve gaps. In equation 3, the length constant $\beta = \frac{d}{\alpha}$. At $L = \beta$ the weighted exposure $y(L) = 0.63 * H$, in other words the exposure is 63% of the peak exposure.

Our analysis to determine the minimum collimation at which the detector behaves like a point detector gives Equation 5, L=5.49d, for the 100 mm pencil chamber. This is derived from our analysis of plots in Figure 3. This means that collimation (L) must be increased to 5.49 times the dimension (d) of the detector before steady state equilibrium is reached according to our 2% cutoff criteria. This number can be interpreted to represent a cutoff above which the detector behaves like a point detector. For a Farmer chamber, this minimum cutoff occurs at 129 mm collimation according to our data.

Several more results are shown in Table 1 above for sleeve gaps used in this experiment. The fitting parameter $\alpha$ is consistent at sleeve gaps greater than 20 mm. For the 10 mm and 20 mm sleeve gaps, $\alpha$ is different possibly due to inaccuracies in collecting the exposure at the center and peripheral regions. A sleeve gap of 100 mm is a pencil chamber with no lead sleeves used. This can be interpreted to mean if a CT scanner has a collimation of >549 mm, the pencil chamber can be used as if it was a point chamber to measure peak dose. Looking at Equation 5 from another perspective, given a certain collimation, we can predict the largest detector size that will act as a point for measuring the peak dose at that collimation. So, for a 160 mm collimation, the largest detector that can act as a point detector is d=L/5.49, or 29 mm long, for convergence at 98%. This notion was also suggested in a previous article [5]. When it comes to predicting the largest collimations at which a given detector will act as a point we can accept that such large collimations are unrealistic. However, we are led to this by the mathematics of the situation and are therefore justified.

Figure 2 shows experimental data of exposure at different sleeve gaps for a fixed collimation. After data collection at the first collimation, the collimation was changed and the data was again collected at different sleeve gaps. The data is fitted by the model in Equation 4. Here, the smallest collimation of 5 mm shows the least variation in exposure according to the data plotted in Figure 2. At 40 mm collimation, the data plotted grows exponentially to about the 60 mm sleeve gap, and then shows some leveling off at the 100 mm sleeve gap. At collimations of 80, 120, 140 and 160 mm, the exposure continues to grow and shows no asymptote. In equation 4, the length constant $\beta = \frac{L}{\alpha}$. At $d = \beta$ the weighted exposure $y(d) = 0.63 * H$, in other words the exposure is 63% of the peak exposure. At $L = 5\beta$, the exposures $y(L)$ and $y(d)$ are 99% of peak exposure H for the cases represented in equation 3 and 4 respectively.

Our analysis of the minimum detector dimension needed at each collimation leads to Equation 6, d=4.94L. This result is derived from data plotted in Figure 4 and implies that if the detector is 4.94 times longer than the collimation, the detector will integrate 98% of the signal in an infinite phantom. This result can also be used to predict minimum phantom lengths needed at different collimations or beam widths. Assuming that the detector spans the phantom in one piece, or in segmented pieces of length equal to the one piece, for integration of the signal, we can set the minimum phantom length for capturing the entire (98%) dose profile at a given collimation. The constant 4.94 is consistent for all collimations except the 5 mm collimation. This is likely due to larger measurement errors at such a small collimation, perhaps brought on by the nature of the sleeve gaps used. If we consider the cutoff to be 1% instead of 2%, then we get d=5.83L. For this relation, at 138 mm collimation we can predict a minimum phantom length of 804.5mm. This number is close to the actual phantom length of 900 mm used by Mori [7], where the dose profiles at both the center and periphery are shown to converge well before the 900 mm phantom ends are reached. At the 2% cutoff a collimation of 160 mm requires a minimum phantom length of 790 mm. This is close to the 750 mm phantom length used in [14]. Looking from a different perspective, equation 6 is also useful for predicting the largest collimation that will allow a given detector to integrate 98% of the signal. We can do this by setting L=d/4.94. So, given a detector of 300 mm active length, the largest collimation that will allow for integrating 98% of the signal should be ~61mm. Of course, this assumes that the phantom is at least as long as the length of the detector.

Since this work was done with a pencil chamber, a criticism may be that it will not apply to another chamber, for example the Farmer chamber. We believe this work will also apply to any type of chamber. This will of course depend on some qualifying factors. Assuming that issues of geometric efficiency, sensitivity, and collection volumes could be normalized, we expect that a sleeve gap of 23.5 mm on a pencil chamber will behave like a farmer chamber of the same length.

In fitting data in Figures 1 and 2 to the model equations an $H_{min}$ different from zero was necessary, suggesting that there was some leakage or penetration around the edges of the lead sleeves.

The overall model is evaluated by using a linear regression analysis of actual measured values and modeled results. Except for measurements done at 5 mm collimation, there is good agreement between model vs. actual measurements as evidenced by the $R^2$ values in Table 3. RMSE is equivalent to the standard deviations of the means of the actual values. This also shows good results for a model that describes the actual measurements.

This work shows that when a detector is used to measure exposure in CT, the exposure measured depends on both the detector dimension and the collimation set. We have shown this both by a mathematical model and supporting experimental data. Our work is different from other works done in this area and is in disagreement with the models shown in those works. We believe our model is accurate due to its ability to predict previously reported phantom lengths accurately. Another advantage is that the model is not extracted from a data fitting approach but is rather developed from an ODE and then used to fit the experimental data. This allows us to set beforehand the variables of detector dimension and collimation and to show step by step how these incorporate into the final model used for fitting the data.

In previous work [6] where the peak exposure measured approached equilibrium at different beam widths/collimations, the approach to equilibrium function contained the beam width as a variable. This is in agreement with our model where exposure is modeled as a function of collimation at fixed detector dimension. However, in these models a parameter d in the exponent was defined as the "width of the broad scatter LSF" [6] (d~100 mm). In addition, the constant term multiplying the term in parenthesis was defined as the scatter to primary (S/P) ratio η. This term was said to not depend on the beam width "a", as defined in those papers, since it was an "impulse-response amplitude" [6]. Other subsequent articles have also defined a parameter d in the exponential term in the same way [15]. Our analysis is different. The model proposed shows that the quantities multiplying the term in parenthesis depends on the sleeve gap, as well as being part of an asymptotic term as defined above in Equation 1. Also, the exponential term is shown to be a function of the beam width or collimation at a given detector dimension d when it pertains to the first question considered above. We have thus defined the term d in the exponent as a detector dimension in mm and shown that it is consistent with experimental data collected by simulation using the approach of leaded sleeve gaps on a pencil chamber. It is noted that in these articles scan length was used. Scan length is related to the collimation set and so our use of collimation is not in conflict with the use of scan length. Obviously this work is also extendable to the stationary cone beam CT or in any scenario where peak exposures are to be measured in an "infinite" phantom.

This work is important because it develops a model based on the solution of a second order linear ODE or first order ODE. Both ODEs were shown to be equivalent and leading to equivalent solutions for the approach to equilibrium function under the stated initial conditions. Our solution reveals important relationships between collimation and detector dimension in the measurement of CT peak exposure. The model is shown to fit the data from experiments carried out with lead sleeves around a pencil chamber. This approach of using lead sleeves wrapped around a pencil chamber as a surrogate for detector dimension is unique in the literature. It is both innovative and simple and allows a simulation of the approach of exposure to an equilibrium value for different collimations and simulated detector sizes. The model solution to the ODEs also is different in that it correctly identifies the detector dimension in the exponential term and also shows how the factors outside the terms in parenthesis are determined.

# CONCLUSION

We present a robust model based on the use of ODEs to describe an approach to equilibrium of weighted peak exposures when a detector is used to measure CT exposure at different collimations. The model fits the data for measurements taken in a 60 cm long body phantom for a 256-slice GE MDCT scanner. We expect that this result can be transferrable to other similar wide beam MDCT scanners. The approach to equilibrium function has been shown to depend on the sleeve gap, a surrogate to the detector size. Approach to equilibrium has also been shown to depend on the collimation used. Minimum detector dimensions are calculated from the data to show the length of integrating detector needed to collect 98% of the exposure at different collimations. In parallel to this, different minimum collimations at which a given detector behaves as a point detector has been calculated and presented. We can ultimately conclude that for peak exposure measurement, the collimation L should be at least 5.49 times longer than the detector dimension d. For integral exposure measurements, the detector dimension d should be at least 4.95 times longer than the collimation L. Finally, we have proven a theorem that says that peak exposures at any collimiation should be measured with detectors whose dimensions approach zero.

# ACKNOWLEDGEMENTS

I would like to thank Dr. Bryan Yoder, Dr. Michael Wayson, and Dr Angela Bruner for reviewing the manuscript and offering useful and insightful comments. I would like to thank Dr. Ian Hamilton for lending me two body phantoms used in this study.